\documentclass[prd,superscriptaddress,unsortedaddress,twocolumn,showpacs,preprintnumbers,amsmath,amssymb]{revtex4}

\usepackage[dvips]{graphicx}
\usepackage{amsmath,amssymb,times}
\usepackage{color}
\usepackage{ulem}

\def\fun#1#2{\lower3.6pt\vbox{\baselineskip0pt\lineskip.9pt
\ialign{$\mathsurround=0pt#1\hfil##\hfil$\crcr#2\crcr\sim\crcr}}}

\newcommand{\beq}{\begin{equation}}
\newcommand{\eeq}{\end{equation}}
\newcommand{\bea}{\begin{eqnarray}}
\newcommand{\eea}{\end{eqnarray}}

\DeclareSymbolFont{boldletters}{OML}{cmm} {b}{it}
\DeclareSymbolFontAlphabet{\mathbit}{boldletters}
\DeclareMathSymbol{\alpha}{\mathalpha}{letters}{"0B}
\DeclareMathSymbol{\beta}{\mathalpha}{letters}{"0C}
\DeclareMathSymbol{\gamma}{\mathalpha}{letters}{"0D}
\DeclareMathSymbol{\delta}{\mathalpha}{letters}{"0E}
\DeclareMathSymbol{\epsilon}{\mathalpha}{letters}{"0F}
\DeclareMathSymbol{\zeta}{\mathalpha}{letters}{"10}
\DeclareMathSymbol{\eta}{\mathalpha}{letters}{"11}
\DeclareMathSymbol{\theta}{\mathalpha}{letters}{"12}
\DeclareMathSymbol{\iota}{\mathalpha}{letters}{"13}
\DeclareMathSymbol{\kappa}{\mathalpha}{letters}{"14}
\DeclareMathSymbol{\lambda}{\mathalpha}{letters}{"15}
\DeclareMathSymbol{\mu}{\mathalpha}{letters}{"16}
\DeclareMathSymbol{\nu}{\mathalpha}{letters}{"17}
\DeclareMathSymbol{\xi}{\mathalpha}{letters}{"18}
\DeclareMathSymbol{\pi}{\mathalpha}{letters}{"19}
\DeclareMathSymbol{\rho}{\mathalpha}{letters}{"1A}
\DeclareMathSymbol{\sigma}{\mathalpha}{letters}{"1B}
\DeclareMathSymbol{\tau}{\mathalpha}{letters}{"1C}
\DeclareMathSymbol{\upsilon}{\mathalpha}{letters}{"1D}
\DeclareMathSymbol{\phi}{\mathalpha}{letters}{"1E}
\DeclareMathSymbol{\chi}{\mathalpha}{letters}{"1F}
\DeclareMathSymbol{\psi}{\mathalpha}{letters}{"20}
\DeclareMathSymbol{\omega}{\mathalpha}{letters}{"21}
\DeclareMathSymbol{\varepsilon}{\mathalpha}{letters}{"22}
\DeclareMathSymbol{\vartheta}{\mathalpha}{letters}{"23}
\DeclareMathSymbol{\varpi}{\mathalpha}{letters}{"24}
\DeclareMathSymbol{\varrho}{\mathalpha}{letters}{"25}
\DeclareMathSymbol{\varsigma}{\mathalpha}{letters}{"26}
\DeclareMathSymbol{\varphi}{\mathalpha}{letters}{"27}
\DeclareMathSymbol{\Gamma}{\mathalpha}{letters}{"00}
\DeclareMathSymbol{\Delta}{\mathalpha}{letters}{"01}
\DeclareMathSymbol{\Theta}{\mathalpha}{letters}{"02}
\DeclareMathSymbol{\Lambda}{\mathalpha}{letters}{"03}
\DeclareMathSymbol{\Xi}{\mathalpha}{letters}{"04}
\DeclareMathSymbol{\Pi}{\mathalpha}{letters}{"05}
\DeclareMathSymbol{\Sigma}{\mathalpha}{letters}{"06}
\DeclareMathSymbol{\Upsilon}{\mathalpha}{letters}{"07}
\DeclareMathSymbol{\Phi}{\mathalpha}{letters}{"08}
\DeclareMathSymbol{\Psi}{\mathalpha}{letters}{"09}
\DeclareMathSymbol{\Omega}{\mathalpha}{letters}{"0A}

\begin{document}
\title{Effective model approach to  meson screening 
masses at finite temperature}

\author{Masahiro Ishii}
\email[]{ishii@phys.kyushu-u.ac.jp}
\affiliation{Department of Physics, Graduate School of Sciences, Kyushu University,
             Fukuoka 812-8581, Japan}

\author{Takahiro Sasaki}
\email[]{sasaki@phys.kyushu-u.ac.jp}
\affiliation{Department of Physics, Graduate School of Sciences, Kyushu University,
             Fukuoka 812-8581, Japan}

\author{Kouji Kashiwa}
\email[]{kashiwa@ribf.riken.jp}
\affiliation{RIKEN/BNL, Brookhaven, National Laboratory, Upton, New York
11973, USA}

\author{Hiroaki Kouno}
\email[]{kounoh@cc.saga-u.ac.jp}
\affiliation{Department of Physics, Saga University,
             Saga 840-8502, Japan}  

\author{Masanobu Yahiro}
\email[]{yahiro@phys.kyushu-u.ac.jp}
\affiliation{Department of Physics, Graduate School of Sciences, Kyushu University,
             Fukuoka 812-8581, Japan}

\date{\today}

\begin{abstract}
Temperature dependence of pion and sigma-meson screening masses is 
evaluated by the Polyakov-loop extended Nambu--Jona-Lasinio (PNJL) model 
with the entanglement vertex. 
We propose a practical way of calculating meson screening masses 
in the NJL-type effective models. 
The method based on the Pauli-Villars regularization 
solves the well-known difficulty that the evaluation 
of screening masses is not easy in the NJL-type effective models.
The PNJL model with the entanglement vertex and 
the Pauli-Villars regularization well reproduces lattice QCD results on 
temperature dependence of the chiral condensate and the Polyakov loop. 
The method is applied to analyze temperature dependence of 
pion screening masses calculated 
with state-of-the-art lattice simulations with success in reproducing 
the lattice QCD results.  
\end{abstract}

\maketitle

\section{Introduction}
Meson masses are not only fundamental quantities of hadrons but also 
a key to know properties of quantum chromodynamics (QCD) vacuum. 
For example, temperature  ($T$) dependence of 
pion and sigma-meson masses is strongly related 
to chiral symmetry restoration of QCD vacuum. 
Such light mesons play an important role 
in nuclear physics as mediators of the nuclear force. 
$T$ dependence of light meson masses affects 
the equation of state particularly around and above 
the pseudocritical temperature $T_c$ of chiral and deconfinement crossover temperature \cite{Bazavov,Borsanyi}.

Lattice QCD (LQCD) is the first-principle calculation of QCD.  
At finite $T$, meson pole (screening) masses are 
calculated from the exponential decay of temporal (spatial) 
mesonic correlation functions. 
LQCD simulations are more difficult for pole masses 
than for screening masses, since 
the lattice size is smaller in the time direction than 
in the spatial direction. This situation becomes more serious 
as $T$ increases. 
For this reason, meson screening masses were calculated 
in most of the LQCD simulations. Recently, 
a state-of-the-art calculation was done for meson screening masses 
in a wide range of 
$T < 4 T_{c} \approx 800$~MeV~\cite{Cheng:2010fe}.

Constructing the effective model is an approach complementary to 
the first-principle LQCD simulation. 
For example, the phase structure and light meson pole masses 
are extensively investigated at finite $T$  by the  
Nambu--Jona-Lasinio (NJL) model~\cite{NJL1,KHKB} and 
the the Polyakov-loop extended Nambu--Jona-Lasinio (PNJL) model
~\cite{Meisinger,Dumitru,Fukushima1,Ghos,Megias,Ratti1,Ciminale,Ratti2,Rossner,Hansen,Sasaki-C,Schaefer,Kashiwa1}.  
The NJL model treats the chiral symmetry breaking, but not 
the confinement mechanism. 
Meanwhile, the PNJL model is designed \cite{Fukushima1} to treat the confinement mechanism approximately in addition to the chiral symmetry breaking. 
In this sense, the PNJL model is superior to the NJL model. 
In the two-flavor PNJL model the chiral and deconfinement 
transitions do not coincide with each other when the model parameters are set 
to reproduce the realistic transition temperature~\cite{Ratti1}, 
whereas the coincidence is seen in the two-flavor LQCD simulations. 
This problem is solved by introducing the four-quark vertex depending 
on the Polyakov loop~\cite{Sakai_EPNJL,Sasaki_EPNJL}. 
The model with the entangle vertex is called 
the entanglement-PNJL (EPNJL) model. 
The EPNJL model can also reproduce the QCD phase structure at imaginary chemical potential~\cite{D'Elia3,FP2010} and at real isospin chemical potential~\cite{Kogut2} where LQCD is feasible.

The NJL-type effective models are quite practical. 
In fact, meson pole masses have been extensively studied with the models. 
However, only a few trials were made so far for 
the evaluation of meson screening masses 
$M_{\xi,{\rm scr}}$~\cite{Kunihiro,Florkowski}; here $\xi$ means 
a species of mesons. 
The model calculations have essentially two problems. 
One problem is that the NJL-type models are nonrenormalizable and hence 
the regularization is needed in the model calculations. 
The regularization commonly used is the three-dimensional momentum cutoff. 
The momentum cutoff breaks Lorentz invariance and thereby the 
spatial correlation function $\eta_{\xi\xi}(r)$ 
has an unphysical oscillation~\cite{Florkowski}. 
This makes the determination of $M_{\xi,{\rm scr}}$ quite difficult, since 
$M_{\xi,{\rm scr}}$ 
is defined from the exponential decay of $\eta_{\xi\xi}(r)$ 
at large distance ($r$):
\begin{equation}
M_{\xi,{\rm scr}}=-\lim_{r\rightarrow \infty}\frac{d \ln{\eta_{\xi\xi}(r)}}{dr}.
\label{scr-mass}
\end{equation}

\begin{figure}[htbp]
\begin{center}
  \includegraphics[width=0.49\textwidth]{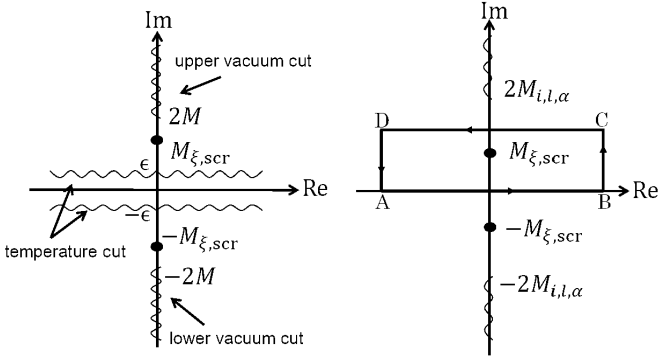}
\end{center}
\caption{Singularities of $\chi_{\xi\xi}(0,\tilde{q})$ 
in the complex-$\tilde{q}$ plane 
based on the previous formulation~\cite{Florkowski} (left)  
and the present formulation 
(right). Cuts are denoted by the wavy lines and poles by the points. 
}
\label{Fig-sing}
\end{figure}

Another problem is the feasibility of numerical calculations. 
In the model approach,  $\eta_{\xi\xi}(r)$ is first obtained 
in the momentum ($\tilde{q}=\pm |\bf q|$) representation 
$\chi_{\xi\xi}(0,\tilde{q}^2)$. 
In the Fourier transformation to the coordinate representation,
\begin{equation}
\eta_{\xi\xi}(r)=\frac{1}{4\pi^{2}ir}\int^{\infty}_{-\infty}d\tilde{q}\hspace{1ex}\tilde{q}\chi_{\xi\xi}(0,\tilde{q}^2)e^{i\tilde{q}r}\hspace{1ex},
\label{chi_r}
\end{equation}
the integrand is slowly damping and highly oscillating particularly 
at large $r$ where $M_{\xi,{\rm scr}}$ is defined. 
This requires heavy numerical calculations. 
It was then proposed that the contour integral was made 
in the complex-$\tilde{q}$ plane~\cite{Florkowski}. 
However, the contour integral is 
still hard to do because of the presence of the temperature cuts 
in the vicinity of the real axis~\cite{Florkowski}; see 
the left panel of Fig.~\ref{Fig-sing}, where note that 
$\epsilon$ is an infinitesimal quantity.

In this paper, we propose 
a practical way of calculating $M_{\xi,{\rm scr}}$ 
in the NJL-type effective models. The first problem is solved by using 
the Pauli-Villars (PV) regularization~\cite{Florkowski,PV} 
that preserves Lorentz symmetry. The EPNJL model 
with the PV regularization well reproduces two-flavor LQCD results 
on $T$ dependence of the chiral condensate and the Polyakov loop. 
The second problem is solved by deriving a new expression 
for $\chi_{\xi\xi}(0,\tilde{q}^2)$. In the expression, 
the contributions of the vacuum and 
temperature cuts to $\eta_{\xi\xi}(r)$ 
are partially canceled in the complex-$\tilde{q}$ plane. 
A pole is well isolated from the resultant cut; 
see the right panel of Fig.~\ref{Fig-sing}. 
The screening mass can therefore be obtained from the location of the pole 
without making the Fourier transform to the coordinate representation. 
The proposed method is applied to analyze $T$ dependence of 
pion screening mass obtained by 
state-of-the-art 2+1 flavor LQCD simulations~\cite{Cheng:2010fe}.

\section{Formalism}
\label{Formalism}
We first recapitulate  the EPNJL model~\cite{Sakai_EPNJL,Sasaki_EPNJL} and derive the equations for 
meson pole and screening masses from the Schwinger-Dyson equation for 
the quark-antiquark scattering.


The Lagrangian density of the two-flavor isospin symmetric EPNJL model is 
defined as 
\begin{align}
 {\cal L}  
=& {\bar q}(i \gamma_\nu D^\nu -m_0)q  + G_{\rm s}(\Phi)[({\bar q}q )^2 
  +({\bar q }i\gamma_5 {\vec \tau}q )^2]
\nonumber\\
 & -{\cal U}(\Phi [A],{\bar \Phi} [A],T) 
\label{L}
\end{align}
with the quark field $q$, the current quark mass $m_0$ and 
the isospin matrix ${\vec \tau}$. 
The coupling constant $G_{\rm s}(\Phi)$ of the 
four-quark interaction depends on the Polyakov loop $\Phi$ as 
\begin{equation}
G_{\rm s}(\Phi)=G_{\rm s}\left[1-\alpha_1\Phi{\bar \Phi} -\alpha_2\left(\Phi^3 + {\bar \Phi}^{3}\right)\right],
\label{EPNJL}
\end{equation}
where 
$D^\nu=\partial^\nu + iA^\nu$ with $A^\nu=\delta^{\nu}_{0}g(A^0)_a{\lambda_a/2}
=-\delta^{\nu}_{0}ig(A_{4})_a{\lambda_a/2}$ for the gauge field $A^\nu_a$, the Gell-Mann matrix $\lambda_a$ and the gauge coupling $g$. 
When $\alpha_1=\alpha_2=0$, the EPNJL model is 
reduced to the PNJL model\cite{Meisinger,Dumitru,Fukushima1,Ghos,Megias,Ratti1,Ciminale,Ratti2,Rossner,Hansen,Sasaki-C,Schaefer,Kashiwa1}.

In the EPNJL model, only the time component of $A_\mu$ is treated as a
homogeneous and static background field, which is governed by the
Polyakov-loop potential~~$\mathcal{U}$. The Polyakov loop $\Phi$ and 
its conjugate ${\bar \Phi}$ are then obtained in the Polyakov gauge by  
\begin{align}
\Phi &= {1\over{3}}{\rm tr}_{\rm c}(L),
~~~~~{\bar \Phi} ={1\over{3}}{\rm tr}_{\rm c}({L^*})
\label{Polyakov}
\end{align}
with $L= \exp[i A_4/T]=\exp[i {\rm diag}(A_4^{11},A_4^{22},A_4^{33})/T]$ 
for the classical variables $A_4^{ii}$ 
satisfying that $A_4^{11}+A_4^{22}+A_4^{33}=0$. 
In the determination of the $A_4^{ii}$ from $\Phi$ and ${\bar \Phi}$,
there is an arbitrariness coming from color symmetry. The arbitrariness
does not change any physics. For zero chemical potential ($\mu=0$), $\Phi$ equals to ${\bar \Phi}$. Hence it is possible to choice $A^{33}_4=0$ and determine the others as 
$A^{22}_4=-A^{11}_4={\rm cos}^{-1}(\frac{3\Phi -1}{2})T$.

 We use the logarithm-type Polyakov-loop potential $\mathcal{U}$ of Ref.~\cite{Rossner}. The parameter set in $\mathcal{U}$ is fitted to reproduce LQCD data at finite $T$ in the pure gauge limit. 
The $\mathcal{U}$ yields the first-order deconfinement phase transition 
at $T=T_0$. In the pure gauge limit, 
LQCD data show the phase transition at $T=270$~MeV.  Hence the parameter 
$T_0$ is often set to $270$~MeV, but the EPNJL model with this value of $T_0$ 
yields a larger value of $T_\mathrm{c}$ for the deconfinement transition 
than the two-flavor LQCD prediction $T^{\rm 2f}_{\rm c}\approx 173\pm 8$~MeV~\cite{Borsanyi,Soeldner,Kanaya}. This problem can be solved by rescaling $T_0$. 
In fact, the EPNJL model with $T_0=200$~MeV and $\alpha_1=\alpha_2=0.20$ 
reproduces the two-flavor-LQCD result.

Making the mean field approximation(MFA) to \eqref{L} leads to 
the linearized Lagrangian density 
\begin{align}
 {\cal L}^{\rm MFA}  
= {\bar q}S^{-1}q  - G_{\rm s}(\Phi)\sigma^2  - 
{\cal U}(\Phi [A],{\bar \Phi} [A],T) 
\label{linear-L}
\end{align} 
with the quark propagator 
\bea
S=\frac{1}{i \gamma_\nu \partial^\nu - i\gamma_0A_4 -M} 
\eea
with the effective quark mass $M=m_0-2G_{\rm s}(\Phi)\sigma$. 
Making the path integral over the quark field, one can get 
the thermodynamic potential (per unit volume) as
\begin{align}
&\Omega_{\rm PNJL}\nonumber\\ 
&= U_{\rm M}+{\cal U}-2 N_{\rm f} \int \frac{d^3 p}{(2\pi)^3}
   \Bigl[ 3 E_p \notag \\
&+ \frac{1}{\beta}
           \ln~ [1 + 3(\Phi+{\bar \Phi} e^{-\beta (E_p-\mu )}) 
           e^{-\beta (E_p-\mu)}+ e^{-3\beta (E_p-\mu)}] \notag\\
&+ \frac{1}{\beta} 
           \ln~ [1 + 3({\bar \Phi}+{\Phi e^{-\beta (E_p+\mu)}}) 
              e^{-\beta (E_p+\mu)}+ e^{-3\beta (E_p+\mu)}]
	      \Bigl]
	      \nonumber\\
\label{PNJL-Omega}
\end{align}
with
$\beta=1/T$, $E_p=\sqrt{{\bf p}^2+M^2}$ and  $U_{\rm M}= G_{\rm s}(\Phi)\sigma^2$, 
where $N_{\rm f}$ is the number of flavors.

Since the momentum integral of (\ref{PNJL-Omega}) diverges, 
we use the PV regularization~\cite{Florkowski,PV}. 
In the scheme, the integral $I(M,q)$ is regularized as 
\begin{eqnarray}
I^{\rm reg}(M,q)=\sum_{\alpha=0}^2 C_\alpha I(M_\alpha,q) ,
\label{PV}
\end{eqnarray}
where $M_0=M$ and $M_\alpha~(\alpha\ge 1)$ are masses of auxiliary 
particles. The parameters $M_\alpha$ and $C_\alpha$ 
should satisfy the condition  
$\sum_{\alpha=0}^2C_\alpha=\sum_{\alpha=0}^2 C_\alpha M_\alpha^2=0$.
We then assume $(C_0,C_1,C_2)=(1,1,-2)$ and $(M_1^2,M_2^2)=(M^2+2\Lambda^2,M^2+\Lambda^2)$. We keep the parameter $\Lambda$ finite 
even after the subtraction \eqref{PV}, since the present model is 
nonrenormalizable. 
The parameters taken are 
$m_0=6.3$ MeV, $G_{\rm s}=5.0$ GeV$^{-2}$ and $\Lambda =0.768$ GeV. 
This parameter set reproduces the pion decay constant $f_{\pi}=93.3$ MeV and 
the pion mass $M_{\pi}=138$ MeV at vacuum. 


We derive the equations for pion and sigma-meson masses, 
following Ref~\cite{Hansen}. Now we consider the case of $\mu=0$. 
The pseudoscalar isovector current with the same quantum number 
as pion is 
\beq
  {J_P}^a(x) = \bar q(x) i \gamma_5 \tau^a q(x)\quad 
\eeq
and the scalar isoscalar current with the same quantum number 
as sigma meson is 
\beq
  {J_S}(x) = \bar q(x) q(x) - \langle\bar q(x) q(x) \rangle .
\eeq
The Fourier transform of the mesonic correlation function 
$\eta_{\xi\xi} (x) \equiv \langle 0 | T \left( J_\xi(x) J^{\dagger}_{\xi}(0) \right) | 0 \rangle$ is 
\begin{align}
\chi_{\xi\xi} (q^2) 
=  i \int d^4x e^{i q\cdot x}
 \langle 0 | {\rm T} \left( J_\xi(x) J^{\dagger}_{\xi}(0) \right)  
 | 0 \rangle  ,
\end{align}
where $\xi=P^a$ for pion and $S$ for sigma meson and ${\rm T}$ stands for 
the time-ordered product. 
Since we deal with only pion and sigma meson, there is no 
mixing term $\chi_{\xi\xi^\prime} ~(\xi^\prime \neq \xi)$. 
Using the random-phase (ring) approximation,
one can obtain the Schwinger-Dyson equation　 
\bea
  \chi_{\xi\xi}(q^2) &=& \Pi_{\xi\xi}(q^2) +2G_{\rm s}(\Phi) \Pi_{\xi\xi} (q^2) \chi_{\xi\xi}(q^2) 
  \label{SD-eq}
\eea
for $\chi_{\xi\xi}(q^2)$, where the one-loop polarization function 
$\Pi_{\xi\xi}$ is defined as 
\bea
  \Pi_{\xi\xi} & \equiv & (-i) \int \frac{d^4 p}{(2\pi)^4} 
  {\rm Tr} \left( \Gamma_\xi iS(p'+q) \Gamma_\xi iS(p') \right) 
\eea
with $p'=(p_{0}+iA_4,{\bf p})$, 
the quark propagator $S(q)$ in the Hartree approximation and 
$\Gamma_\xi=\Gamma^a_{P}=i\gamma_5 \tau^a$ for pion and 
$\Gamma_\xi=\Gamma_{S}= 1$ for sigma meson. 
The solution to \eqref{SD-eq} is 
\bea
\chi_{\xi\xi} &=& \frac{\Pi_{\xi\xi}(q^2)}{1 - 2G_{\rm s}(\Phi) \Pi_{\xi\xi}(q^2)} . 
\eea
At $T=0$, $\chi_{\xi\xi}$ and $\Pi_{\xi\xi}$ are functions 
of $q^2=q_0^2-{\bf q}^{2}$, but for later convenience we denote them 
as $\chi_{\xi\xi}(q_0^2,{\bf q}^2)$ and $\Pi_{\xi\xi}(q_0^2,{\bf q}^2)$.  
For $T=0$, $\Pi_{\xi\xi}$ is explicitly obtained by 
\begin{eqnarray}
\Pi_{SS}&=&i\int {d^4p\over{(2\pi )^4}}{\rm Tr}\Bigl[{\{\gamma_\mu (p'+q)^\mu +M\}(\gamma_\nu p'^\nu +M)\over{\{(p'+q)^2-M^2\}(p'^2-M^2)}}\Bigr]
\nonumber\\
&=&2iN_{\rm f}[I_1+I_2-(q^2-4M^2)I_3], 
\label{Pi_SS}
\\
\Pi_{PP}&=&i\int {d^4p\over{(2\pi )^4}}{\rm Tr}\Bigl[(i\gamma_5\tau^a){\{\gamma_\mu (p'+q)^\mu +M\}\over{\{(p'+q)^2-M^2\}}}
\nonumber\\
&&\times (i\gamma_5\tau^a)
{(\gamma_\nu p'^\nu +M)\over{(p'^2-M^2)}}\Bigr]
\nonumber\\
&=&2iN_{\rm f}[I_1+I_2-q^2I_3], 
\label{Pi_PP}
\end{eqnarray}
with 
\begin{eqnarray}
I_1&=&\int {d^4p\over{(2\pi )^4}}{\rm tr_c}\Bigl[{1\over{p'^2-M^2}}\Bigr],
\label{I1}
\\
I_2(q_0^2,{\bf q}^2)&=&\int {d^4p\over{(2\pi )^4}}{\rm tr_c}\Bigl[{1\over{(p'+q)^2-M^2}}\Bigr],
\label{I2}
\\
I_3(q_0^2,{\bf q}^2)&=&\int {d^4p\over{(2\pi )^4}}{\rm tr_c}\Bigl[{1\over{\{(p'+q)^2-M^2\}(p'^2-M^2)}}\Bigr] ,
\nonumber\\
\label{I3}
\end{eqnarray}
where ${\rm tr}_{\rm c}$ means the trace of color matrix. For finite $T$, the corresponding equations are 
obtained by the replacement
\begin{align}
&p_0 \to i \omega_l = i(2l+1) \pi T, 
\nonumber\\
&\int \frac{d^4p}{(2 \pi)^4} 
\to iT\sum_{l=-\infty}^{\infty} \int \frac{d^3p}{(2 \pi)^3}. 
\label{finte_T_mu}
\end{align}

The meson pole mass $M_{\xi}$ is a pole of 
$\chi_{\xi\xi}(q_0^2,{\bf q}^2)$. Taking the rest frame $q=(q_0,{\bf 0})$ 
for convenience, one can get the equation for $M_{\xi}$ as 
\begin{align}
\big[1 - 2G_{\xi\xi} \Pi_{\xi\xi}(q_0^2,0)\big]\big|_{q_0=M_\xi}=0.   
\label{mmf}
\end{align}
The method of calculating meson pole masses is well established 
in the PNJL model~\cite{Hansen}.

The meson screening mass $M_{\xi,{\rm scr}}$ defined with \eqref{scr-mass} 
is obtained by making the Fourier transform of 
$\chi_{\xi\xi} (0,\tilde{q}^{2})$ as shown in \eqref{chi_r}. 
In the previous formalism~\cite{Florkowski}, however, the procedure 
requires heavy numerical calculations in the $I_3^{\rm reg}$ part, 
as shown below, where $I_3^{\rm reg}$ means a function 
after the PV regularization. 
Taking the $l$ summation before the $p$ integral in \eqref{finte_T_mu}, 
one can describe 
$I_3^{\rm reg}(0,\tilde{q}^{2})$ as 
the sum of the vacuum and temperature parts, 
$I_{3,{\rm vac}}^{\rm reg}$ and $I_{3,{\rm tem}}^{\rm reg}$, defined by 
\begin{eqnarray}
\hspace{-3ex}I_{3,{\rm vac}}^{\rm reg}(0,\tilde{q}^{2})
&=&\frac{-iN_c}{16\pi^2}\sum_{\alpha=0}^{2}C_\alpha\left[\ln{M_\alpha^2}+f_{\rm vac}\left(\frac{2M_{\alpha}}{\tilde{q}}\right)\right] ,~~~
\\
f_{\rm vac}(x)&=&\sqrt{1+x^2}\ln{\left(\frac{\sqrt{1+x^2}+1}{\sqrt{1+x^2}-1}\right)} 
\end{eqnarray}
and
\begin{eqnarray}
\hspace{-2ex}I_{3,{\rm tem}}^{\rm reg}(0,\tilde{q}^{2})
&&\hspace{-4ex}=\frac{iN_c}{16\pi^2}\sum_{\alpha=0}^{2}C_\alpha\int_0^\infty d\tilde{p}~f_{\rm tem}(\tilde{p},\tilde{q})\left(F_{\tilde{p}}^{-}+F_{\tilde{p}}^{+}\right),~~~~~~
\\
f_{\rm tem}(\tilde{p},\tilde{q})&=&\frac{1}{E_{\tilde{p}}}\frac{\tilde{p}}{\tilde{q}}\ln{\left(\frac{(\tilde{q}-2\tilde{p})^{2}+\epsilon^{2}}{(\tilde{q}+2\tilde{p})^{2}+\epsilon^{2}}\right)} , 
\label{f-tem}
\end{eqnarray}
where the Fermi distribution functions $F_{\pm}$ are defined as 
\begin{eqnarray}
F^{\pm}_{\tilde{p}}=F^{\pm}(\tilde{p},A_4,T)&=&\frac{1}{N_c}\sum_{i=1}^{N_c}{1\over{e^{(E_{\tilde{p}} \pm 
iA^{ii}_4 )/T}+1}} . 
\end{eqnarray}
In \eqref{f-tem}, the $\epsilon^2$ term is added 
to make the $\tilde{p}$ integral well defined at $\tilde{q}=\pm2\tilde{p}$, 
but this requires the limit of $\epsilon \to 0$.

As shown in the left panel of Fig.~\ref{Fig-sing}, 
$f_{\rm vac}({2M_{\alpha}}/{\tilde{q}})$ and 
$f_{\rm tem}(\tilde{p},\tilde{q})$ have the vacuum and temperature cuts in 
the complex $\tilde{q}$ plane, respectively. 
In \eqref{chi_r}, the cuts contribute to 
the $\tilde{q}$ integral in addition to 
the pole at $\tilde{q}=iM_{\xi,{\rm scr}}$ defined by
\begin{align}
\big[1 - 2G_{\xi\xi} \Pi_{\xi\xi}(0,\tilde{q}^2)\big]\big|_{\tilde{q}=iM_{\xi, {\rm scr}}}=0  .  
\label{meson_screening}
\end{align}
It is not easy to evaluate the temperature-cut contribution, 
since in \eqref{chi_r} the integrand is slowly damping and highly oscillating 
with $\tilde{q}$ near the real axis in the complex $\tilde{q}$ plane. 
Furthermore we have to take the limit of $\epsilon \to 0$ finally.

A hint of solving this problem is in the high-$T$ limit where $G_{S}=0$. 
In this situation, it is known~\cite{Florkowski} 
that the vacuum- and temperature-cut 
contributions partially cancel each other. 
We then extend the discussion to general $T$. Using the formula
\begin{eqnarray}
{1\over{e^x+1}}={1\over{2}}-\sum_{l=-\infty}^\infty {x\over{(2l+1)^2\pi^2+x^2}}, 
\label{formula}
\end{eqnarray}
we can rewrite $I_3^{\rm reg}(0,\tilde{q})$ as
\begin{eqnarray}
&&I_{3,{\rm tem}}^{\rm reg}(0,\tilde{q}^2) \nonumber\\
&=&-I_{3,{\rm vac}}^{\rm reg}(0,\tilde{q}^{2})+iT\sum_{i=1}^{N_c}\sum_{l=-\infty}^\infty\sum_{\alpha=0}^2C_{\alpha}
\nonumber\\
&\times & \int {d^3p\over{(2\pi )^3}}
\Bigl[{1\over{{\bf p}^2+M_{i,l,\alpha}^2}}{1\over{({\bf p}+{\bf q})^2+M_{i,l,\alpha}^2}}\Bigr],
\label{I_3_rewritten}
\end{eqnarray}
where 
\begin{equation}
M_{i,l,\alpha}(T)=\sqrt{M_{\alpha}^2 + \{(2l+1)\pi T+A_{4}^{ii}\}^2 }. 
\label{KK_mode}
\end{equation}
Obviously, the first term in the right-hand side of (\ref{I_3_rewritten}) 
cancels $I_{3,{\rm vac}}^{\rm reg}$ in $I_{3}^{\rm reg}$. 
To maintain this cancellation, we have to introduce 
the same regularization to both $I_{3,{\rm tem}}^{\rm reg}$ 
and $I_{3,{\rm vac}}^{\rm reg}$, 
although $I_{3,{\rm tem}}$ is finite.    
Consequently we get  
\begin{eqnarray}
&&I_{3}^{\rm reg}(0,\tilde{q}^{2})\nonumber\\
&=&{iT\over{2\pi^2}}\sum_{i,l,\alpha}C_{\alpha}\int_0^1dx\int_0^\infty d\tilde{k}{\tilde{k}^2\over{[\tilde{k}^2+(x-x^2)\tilde{q}^2+M_{i,l,\alpha}^2]^2}}
\nonumber\\
&=&{iT\over{4\pi \tilde{q}}}\sum_{i,l,\alpha}C_{\alpha}  \sin^{-1}{\Bigl({{\tilde{q}\over{2}}\over{\sqrt{{\tilde{q}^2\over{4}}+M_{i,l,\alpha}^2}}}\Bigr)}. 
\label{I_3_final}
\end{eqnarray}
We have numerically checked 
that the convergence of $l$ summation is quite fast in (\ref{I_3_final}). 
Each term of $I_3^{\rm reg}(0,\tilde{q})$ 
has only two cuts starting from $\pm 2iM_{i,l,\alpha}$ 
on the imaginary axis in the complex $\tilde{q}$ plane. 
The cuts are shown 
in the right panel of Fig.~\ref{Fig-sing}. 
The lowest branch point is $\tilde{q}=2iM_{i=1,l=0,\alpha=0}$. Hence 
$2M_{i=1,l=0,\alpha=0}$ is regarded as ``threshold mass'' 
in the sense that the meson screening-mass spectrum 
becomes continuous above the point. 

If $M_{\xi,{\rm scr}}<2M_{i=1,l=0,\alpha=0}$, 
the pole at $\tilde{q}=iM_{\xi, {\rm scr}}$ 
is well isolated from the cut. Hence one can take the contour 
(A$\to$B$\to$C$\to$D$\to$A) shown 
in the right panel of Fig.~\ref{Fig-sing}. The $\tilde{q}$ integral 
of $\tilde{q}\chi_{\xi\xi}(0,\tilde{q}^2)e^{i\tilde{q}r}$ 
on the real axis in \eqref{chi_r} is then obtained from the residue 
at the pole and the line integral from point C to point D.
The former behaves as $\exp[-M_{\xi,{\rm scr}} r]/r$ at large $r$ 
and the latter as $\exp[{-2M_{i=1,l=0,\alpha=0}}r]/r$. 
The behavior of $\eta_{\xi\xi}(r)$ at large $r$ is thus 
determined by the pole. 
One can then determine the screening mass from the location of 
the pole in the complex-$\tilde{q}$ plane 
without making the $\tilde{q}$ integral. 
In the high-$T$ limit, the condition tends to $M_{\xi,{\rm scr}}<2 \pi T$.

\section{Numerical Results}



 
For $T$ dependence of the chiral
condensate $\sigma$ and the Polyakov loop $\Phi$ 
in two-flavor LQCD simulations~\cite{Karsch,Kaczmarek}, 
the EPNJL model with the PV regularization yields the same quality 
of agreement with the LQCD data as the model with the 3d-momentum cutoff 
regularization~\cite{Sasaki_EPNJL}.

The pion screening mass $M_{\pi, {\rm scr}}$ obtained by 
state-of-the-art 2+1 flavor LQCD simulations~\cite{Cheng:2010fe}  
is now analyzed by the present two-flavor EPNJL model simply, 
since the meson is composed of $u$ and $d$ quarks. 
This is a quantitative analysis, because the finite lattice-spacing effect is 
not negligible in the simulations. 
The chiral transition temperature evaluated is 
$T^{\rm 3f}_c=196~{\rm MeV}$ 
in the simulations~\cite{Cheng:2010fe}, although it becomes 
$T^{\rm 3f}_c=154\pm9~{\rm MeV}$ 
in finer 2+1-flavor LQCD simulations~\cite{Bazavov} close 
to the continuum limit. 
Therefore, we rescale the LQCD results of Ref.~\cite{Cheng:2010fe} with
multiplying them by the factor $154/196$ to reproduce 
$T^{\rm 3f}_c=154\pm9~{\rm MeV}$. 
The model parameters, $m_0$ and 
$T_{0}$, are refitted to reproduce the rescaled 2+1 flavor LQCD data, i.e., 
$M_{\pi}=175~{\rm MeV}$ at vacuum and $T^{\rm 3f}_c=154\pm9~{\rm MeV}$; 
the resulting values are $m_0=10.3~{\rm MeV}$ and $T_{0}=156~{\rm MeV}$.
The variation of $m_0$ from the original value $6.3$ to 
$10.3~{\rm MeV}$ little changes $\sigma$ and $\Phi$.

\begin{figure}[t]
\begin{center}
   \includegraphics[width=0.49\textwidth]{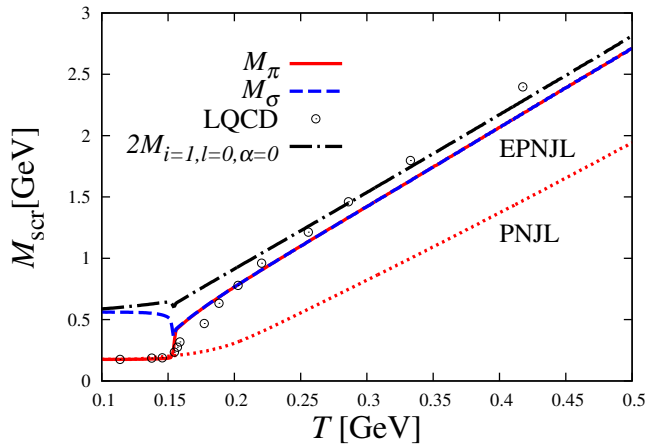}
\end{center}
\caption{$T$ dependence of pion and sigma-meson screening masses, 
$M_{\pi,{\rm scr}}$ and $M_{\sigma,{\rm scr}}$. 
The solid and dashed lines  denote 
$M_{\pi,{\rm scr}}$ and $M_{\sigma,{\rm scr}}$ calculated by 
the EPNJL model with $\alpha_1=\alpha_2=0.31$, respectively, whereas 
the dotted line corresponds to $M_{\pi,{\rm scr}}$ calculated 
with the PNJL model with $\alpha_1=\alpha_2=0$. 
The open circles show $M_{\pi,{\rm scr}}$ obtained by 
2+1 flavor LQCD simulations~\cite{Cheng:2010fe}. 
The dot-dashed line stands for the threshold mass. 
}
\label{Fig-screening}
\end{figure}

As shown in Fig.~\ref{Fig-screening}, the $M_{\pi, {\rm scr}}$ calculated 
with the EPNJL model (solid line) well reproduces the LQCD result 
(open circles), 
when $\alpha_1=\alpha_2=0.31$.
In the PNJL model with $\alpha_1=\alpha_2=0$, 
the model result (dotted line) largely underestimates the LQCD 
result, indicating that the entanglement is important. 
The dashed line denotes the sigma-meson screening mass 
$M_{\sigma, {\rm scr}}$ obtained by the EPNJL model with 
$\alpha_1=\alpha_2=0.31$. 
The solid and dashed lines are lower than the threshold mass 
$2M_{i=1,l=0,\alpha=0}$ (dot-dashed line). This guarantees 
that the $M_{\pi, {\rm scr}}$ and $M_{\sigma, {\rm scr}}$ determined 
from  the location of the single pole in the complex-$\tilde{q}$ plane 
agree with those from the exponential decay 
of $\eta_{\xi\xi}(r)$ at large $r$. 
The chiral restoration takes place at $T=T_c=154$~MeV, since 
$M_{\pi, {\rm scr}}=M_{\sigma, {\rm scr}}$ there. 
After the restoration, the screening masses 
rapidly approach the threshold mass and finally $2 \pi T$. 
The threshold mass is thus an important concept 
to understand $T$ dependence of screening masses.

\section{Summary}

We have proposed a practical way of calculating meson 
screening masses $M_{{\xi,\rm scr}}$ in the NJL-type models. 
This method based on the PV regularization 
solves the well-known difficulty that the evaluation of $M_{{\xi, \rm scr}}$ 
is not easy in the NJL-type effective models. 
In the previous formalism~\cite{Florkowski}, the vacuum and temperature cuts 
appear in the complex-$\tilde{q}$ plane. The contributions 
to the mesonic correlation function are partially canceled 
in the present formalism. The branch point of 
the resultant cut can be regarded as the threshold mass. 
The pion and sigma-meson screening masses 
rapidly approach the threshold mass $2M_{i=1,l=0,\alpha=0}(T)$ 
after the chiral restoration.

\noindent
\begin{acknowledgments}
The authors thank J. Takahashi for useful discussion. T.S. is supported
 by JSPS KAKENHI Grant No. 23-2790. K.K is supported by 
RIKEN Special Postdoctoral Researchers Program.
\end{acknowledgments}


\end{document}